\documentclass{amsproc}

\usepackage{latexsym}
\usepackage{amsfonts}
\usepackage{amsthm}
\usepackage{graphicx}
\usepackage{amsmath}
\usepackage{amssymb}

\newtheorem{theorem}{Theorem}[section]

\theoremstyle{definition}

\theoremstyle{remark}

\numberwithin{equation}{section}

\begin{document}

\title[Publicly Verifiable Secret Sharing Using Non-Abelian groups]
{Publicly Verifiable Secret Sharing Using Non-Abelian Groups}


\author[D. Kahrobaei]{Delaram Kahrobaei}
\address{CUNY Graduate Center, and City Tech, City University of New York}%
\email{DKahrobaei@GC.Cuny.edu}
\thanks{The Research of the first author has been supported by PSC CUNY research foundation as well as City Tech Foundation}

\author[E.Vidaurre]{Elizabeth Vidaurre}
\address{CUNY Graduate Center, City University of New York}%
\email{EVidaurre@GC.Cuny.edu}%
\thanks{The Research of the second author has been supported by NSF LS-AMP}

\subjclass[2000]{Primary }

\date{\today}

\begin{abstract}
In his paper  \cite{MS}, Stadler develops techniques for improving the security of existing secret sharing protocols by allowing to check whether the secret shares given out by the dealer are valid. In particular, the secret sharing is executed over abelian groups. In this paper we develop similar methods over non-abelian groups.
\end{abstract}

\maketitle

\section{Introduction to Publicly Verifiable Secret Sharing}
Secret sharing is the process which involves a dealer and $n$ participants. The dealer picks a secret and hands out to each participant an element, not equal to the secret, called a share through a secure channel. When any $k$ of the participants come together, they can compute the secret, where $k$ is called the threshold. Secret sharing has the property that if any $k-1$ participants come together, it is difficult for them to deduce the secret. The main example of this process is called Shamir's secret sharing scheme \cite{AS}. Stadler uses it in his first example of PVSS. The main application is the situation in which there is a bank with $n$ managers and at least $k$ managers have to be together to open a vault.

The method of secret sharing depends on the benevolence of the dealer because any party involved must trust that the dealer is distributing valid shares to each participant. Verifiable secret sharing adds a layer of security to the scheme by solving the problem of a cheating dealer. In other words, a verifiable secret sharing (VSS) sheme prevents the dealer from distributing a share to a participant that, together with an appropriate number of other shares, does not yield the secret.

The goal of publicly verifiable secret sharing (PVSS) is to allow anyone to verify that the participants received valid shares. In particular, $P_i$ can check that $P_j$ has a valid share. Applications of PVSS are software key escrow and design of electronic cash systems. An example of key escrow is Micali's fair cryptosystems \cite{SM}.

In practice, the protocols proposed use a similar method for accomplishing their respective goals. In a VSS scheme, the dealer would make one or more pieces of information public as proof. Participants would then compute a value using their secret share and compare it to the public proof. In the PVSS scheme, both the dealer and participants publish encrypted values of their secret information. It is preferable that the proof and/or encrypted values involve the least amount of pieces of information possible to prevent the dealer from providing proof that a fake share is valid to a particular participant (defeating the purpose of the scheme).

A VSS scheme can be \textsl{non-interactive}, meaning that the participants are not required to interact with each other in order to verify the validity of their shares. Moreover, in a PVSS scheme, the dealer distributes the shares to each participant using an assymetric key encryption algorithm. Using the public information and possible additional interaction with the dealer, any person can check that the encrypted secret share is valid. In the case that no interaction with the dealer is required, the PVSS scheme is called \textsl{non-interactive}.

This paper describes the two protocols developed by Stadler, both of which rely heavily on the well-known El Gamal encryption sheme. Next, we illustrate a new VSS sheme that uses nonabelian groups. Lastly, we attempt to mimick Stadler's schemes using the non abelian version of El Gamal's scheme, however we were unable to efficiently use all pieces of information, making the scheme insecure.

\subsection{Discrete Logarithm and $\mathbb{Z}_p$ scheme}
The following describes a Shamir's secret sharing scheme with an additional non-interactive VSS and PVSS protocol. In the PVSS protocol, the dealer uses El Gamal's scheme to distribute the shares and then proves to a verifier that the pair $(A,B)$ associated to the participant $P_i$ encrypts the discrete logarithm of a public element $V$. (see  \cite{MS})
\begin{itemize}
\item Fixed: $p$ a large prime, $q=\frac{(p-1)}{2}$ prime, $h \in \mathbb{Z}_p^*$ order $q$, $G$ a group of order $p$, $g$ a generator of G, $s \in \mathbb{Z}$ is the secret, $k$ threshold.
\item Public info: $S=g^s$, nonzero $x_i \in \mathbb{Z}_p$ assigned to $P_i$. 
, $F_j=g^{f_j}$ for random $f_j \in \mathbb{Z}_p$ and $j<k$.
\item Private to $P_i$: 
$s_i=s+\displaystyle\sum\limits_{j=1}^{k-1}f_i x_i^j $ (mod p).
\item Secret can be recovered using Lagrange interpolation.
\item VSS algorithm: $P_i$ computes $S_i=S \displaystyle\prod\limits_{j=1}^{k-1}F_j^{x_i^j}$ and if $S_i=g^{s_i}$, then $P_i$ has a valid share.
\item PVSS algorithm: 
\begin{itemize}
\item $P_i$ choose a secret key $z \in \mathbb{Z}_q$ and publishes $y=h^z \pmod{p}$
\item the element $V=g^v$ of $G$ and the pair $(A, B)=(h^\alpha,v^{-1} y^\alpha)\pmod{p}$ are  made public
\item $P_i$ can retrieve his share by calculating $m=A^zB^{-1} \pmod{p}$
\item for some fixed $l \approx 100$ and $i$ such that $1\leq i \leq l$, dealer/prover chooses $w_i \in \mathbb{Z}_q$ to compute $t_{hi}=h^{w_i} \pmod{p}$ and $t_{gi}=g^{y^{w_i}}$. 
\item Using a cryptographically strong hash-function (for an in-depth discussion description of hash-functions see \cite{JB}), $\mathcal{H}_l:\{0,1\}^* \rightarrow \{ 0,1 \}^l$, she publishes  $$(c_1, \ldots, c_l)=\mathcal{H}_l(V||A||B||t_{h1}||t_{g1}||t_{h2}||t_{g2}|| \ldots ||t_{hl}||t_{gl})$$ He/she also publishes $$(r_1, \ldots, r_l)=(w_1-c_1 \alpha \pmod{q}, \ldots,w_l-c_l \alpha \pmod{q})$$
\item verifier would compute $t_{hi}=h^{r_i}A^{c_i} \pmod{p}$ and $t_{gi}=(g^{1-c_i}V^{c_iB})^{y^{r_i}}$ and then check whether $\mathcal{H}_l(V||A||B||t_{h1}||t_{g1}||t_{h2}||t_{g2}|| \ldots ||t_{hl}||t_{gl})$ is $(c_1, \ldots, c_l)$.
\end{itemize}
\end{itemize}

\subsection{$e$th root and $\mathbb{Z}_n$ scheme}
For this interactive PVSS scheme, the dealer also uses El Gamal's scheme and then must prove that the pair $(A,B)$ encrypts the e-th root of a public element $M$ (see  \cite{MS})
\begin{itemize}
\item secret to $P_i$: random $z \in \mathbb{Z}_n$
\item secret to dealer: random $\alpha \in \mathbb{Z}_n$
\item public: $g \in \mathbb{Z}_n^*$, $y=g^z \pmod{n}$, $(A,B)=(g^\alpha, m y^\alpha)$, $M= m^e$
\item $P_i$ can retrieve his share by calculating $m=A^{-z}B \pmod{n}$ 
\item dealer picks $w \in \{0, \ldots, \lceil 2^ln^{l+\epsilon} \rceil \}$ and makes $t_g=g^w \pmod{n}$ and $t_y=y^{ew}$ public
\item the verifier publishes $c\in \{ 0, \ldots , 2^l-1\}$
\item the dealer publishes $r=w-c\alpha$
\item the verifier checks that $t_G=g^rA^c \pmod{n}$ and $t_y= y^{er}(B^e/M)^c \pmod{n}$
\end{itemize}

\section{New schemes}
In recent years, non-abelian groups have been used in cryptography. One of the first cryptosystems over a non-abelian group was suggested by Anshel-Anshel-Goldfeld \cite{AAG}. Conjugation in non-abelian groups is central to the cryptosystems proposed by \cite{KL}. In particular \cite{HKS} and \cite{DP} proposed new secret sharing protocols using group presentations. Also \cite{KK} non-abelian El Gamal key exchange has been used. For more information on group-based cryptography see \cite{MSS} and \cite{FHKR}
In this paper we are proposing a new PVSS and VSS protocols using non-abelian groups.

\subsection{Non-Commutative Key Exchange using Conjugacy} \cite{KK}

In this section, we discuss the use of conjugation in protocols over non-abelian groups as background to the new protocols proposed. Suppose $G$ is a non-abelian group and $S$, $T \subset G$ such that $[S,T]=1$. Bob takes $s \in S, b \in G$ and publishes $b$ and $c=b^s$ as his
public keys, keeping $s$ as his private key. Here $b^s=s^{-1} b s $. If Alice wishes to
send $x \in G$ as a session key to Bob, she first chooses a random
$t \in T$ and sends $$E=x^{(c^t)}$$ to Bob, along with the header
$$h=b^t.$$ Bob then calculates $(b^t)^s=(b^s)^t=c^t$ with the header.
He can now compute $$E'=(c^t)^{-1}$$ which allows him to decrypt
the session key,
$$(x^{(c^t)})^{E'}=(x^{(c^t)}){^{(c^t)}}^{-1}=x.$$The element $x
\in G$ can now be used as a session key.

The feasibility of this protocol rests on the assumption that
products and inverses of elements of $G$ can be computed
efficiently.  To deduce Bob's private key from public information
would require solving the equation $c=b^s$ for $s$, given the
public values $b$ and $c$. This is called the {\em conjugacy
search problem} for $G$. Thus the security of this scheme rests on
the assumption that there is no fast algorithm for solving the
conjugacy search problem for the group $G$.

\subsection{PVSS using non-abelian groups}
Authentication schemes described in \cite{MSS} use conjugation, which of course require non-abelian groups. Although authentication serves a different purpose, the method also works for PVSS. 

An algorithm analogous to one of Stadler's starts out with the non-abelian El Gamal. Each participant randomly chooses his private key $s \in S$ and publishes $b$ and $c=b^s$. Here $b^s=s^{-1} b s $. The dealer then picks a random $t\in T$ and publishes $(A,B)=(b^t, x^{c^t})$. Consequently, the participant will find that his secret share is $x=B^{(A^s)^{-1}}$. For  verification, the dealer must prove that the pair $(A,B)$ encrypts the element with which a public element $N$ and $n$ are conjugate. The dealer chooses a random $y, w\in G$ and publishes $N=n^x$, $t_h=b^w$ and $t_g=b^{y^w}$. The verifier publishes $r\in \{0,1\}$. If $r=0$, then the dealer sends $c=wt$. If $r=1$, then the dealer sends $c=w$. Then the verifier can check that $t_h=A^c$

\subsection{VSS using non-abelian groups}

Suppose there are $n$ participants and each is given a secret share so that at least $t=n-1$ of them have to be together to obtain the secret $s$. Let $G$ be a nonabelian group where the search conjugacy problem is hard and $F$ be an abelian subset with $n$ elements. The dealer secretly sends $f_i$ to each participant $P_i$. Next, for every $i\leq n$, the following are published $$S=(\prod_{i=1}^n{f_i})^{-1}s\prod_{i=1}^n{f_i} \hspace{1.5cm} \text{and} \hspace{1.5cm} h_i = (\prod_{i \neq j}{f_j})^{-1}s\prod_{i \neq j}{f_j} .$$ Any $t$ participants can recover the secret by conjugating $h_i$ by the inverse of the product of their shares, where $i$ is the missing participant. In order for $P_i$ to verify that his/her share is valid, s/he can check that $f_i^{-1} h_i f_i = S$. Lastly, if $P_i$ and $P_j$ want to verify that each other's shares are valid, then they can check that $f_i^{-1} h_j f_i = f_j^{-1} h_i f_i$ without making their secret shares known to the other participant.

Clearly, the platform group cannot be abelian as conjugation is heavily used. If the group is given by a presentation, then the elements in the subset $F$ can be any elements that have their (pairwise) commutators in the presentation of the group. If there are not enough of these elements, then powers of any one of these elements can serve as another secret share; the only problem with this is that the scheme becomes less secure in this case. Examples of non-abelian groups that can be used are polycyclic and metabelian groups. Metabelian groups would be particularly convenient as a platform group because it would be easy to find commuting elements.

Alternatively, defining $h_i = (\prod_{j \in H_i}{f_j})^{-1}s\prod_{ j \in H_i}{f_j} $ where $H_i$ is a subset of $F$ with $t$ elements allows for any threshold $t$. Similarly, any $t$ participants can recover the secret by conjugating the appropriate $h_i$ by the inverse of the product of their shares. However, the dealer has not published enough information for a participant to verfify that his share is vaild. 

The requirement that the search conjugacy problem be hard in the platform group is necesarry for the security of the scheme. If the search conjugacy problem were efficiently solvable in the group, then an adversary could determine $f_i$ from $S$ and $h_i$ and therefore recover the secret.


\bibliographystyle{amsplain}

\end{document}